\documentclass[prl,aps,twocolumn,showpacs,floats,amssymb,amsfonts]{revtex4}
\usepackage{graphicx}
\usepackage{bm}

\begin{document}

\title{
Fluctuation-dissipation ratio of a spin glass in the aging regime
}

\author{
D. H\'{e}risson and M. Ocio }

\affiliation{DSM/ Service de Physique de l'\'{E}tat Condens\'{e}, CEA-Saclay, 91191 Gif sur Yvette Cedex,
France.}

\begin{abstract}
We present the first experimental determination of the time
autocorrelation $C(t',t)$ of magnetization in the non-stationary
regime of a spin glass. Quantitative comparison with the response,
the magnetic susceptibility $\chi(t',t)$, is made using a new
experimental setup allowing both measurements in the same
conditions. Clearly, we observe a non-linear
fluctuation-dissipation relation between $C$ and $\chi$, depending
weakly on the waiting time $t'$. Following theoretical
developments on mean-field models, and lately on short range
models, it is predicted that in the limit of long times, the
$\chi(C)$ relationship should become independent on $t'$. A
scaling procedure allows us to extrapolate to the limit of long
waiting times.
\end{abstract}

\maketitle

Almost half a century ago, derivation of the fluctuation
dissipation theorem (FDT) \cite{Callen,Kubo} which links the
response function of a system to its time autocorrelation
function, made it possible to work out dynamics from the knowledge
of statistical properties at equilibrium. Nevertheless, this
progress was limited by severe restrictions. FDT applies only to
ergodic systems at equilibrium. Yet, such systems represent a very
limited part of natural objects, and there is now a growing
interest on non-ergodic systems and on the related challenging
problem of the existence of fluctuation dissipation (FD) relations
valid in off-equilibrium situations.

A way to extend equilibrium concepts to non-equilibrium situations
is to consider systems in which single time dependent quantities
(like the average energy) are near equilibrium values though
quantities which depend on two times (like the response to a
field) are not. Spin glasses \cite{Binder} are such systems. They
remain strongly non-stationary even when their rate of energy
decrease has reached undetectable values. In the absence of any
external driving force, they slowly evolve towards equilibrium,
but never reach it, even on geological time-scales. In these
conditions, FDT is not expected to hold. A quite general FD
relation can be written as \cite{CuKu2,CuKu3} $R(t',t)=\beta
X(t',t)\partial C(t',t)/\partial t'$, where $R(t',t)$ is the
impulse response of an observable to its conjugate field,
$C(t',t)$ the autocorrelation function of the observable and
$\beta={1/k_B T}$. FDT corresponds to $X=1$. Determination of $X$,
the fluctuation-dissipation ratio (FDR), or of an ``effective
temperature", $T_{eff}=T/X$, is the aim of many recent theoretical
studies which predicted a generalization of FDT
\cite{CuKu1,CuKu2,CuKu3} in ``weak ergodicity breaking" systems
\cite{Bouchaud}. In the asymptotic limit of large times, it is
conjectured that the FDR should depend on time only through the
correlation function: $X(t',t)=X(C(t',t))$ for $t'$ (and $t>t'$)
$\rightarrow\infty$. The dependence of $X$ on $C$ would reflect
the level of thermalization of different degrees of freedom within
different timescales \cite{CuKu3}. Thus, the integrated forms of
the FD relation would become
$\chi(t',t)=\beta\int_{C(t',t)}^{C(t,t)}X(C)dC$ (susceptibility
function) and $\sigma(t',t)=\beta\int_0^{C(t',t)}X(C)dC$
(relaxation function). They would depend on $t$ and $t'$ only
through the value of $C$. The field cooled magnetization would
read $ \chi _{FC} = \beta \int_0^{C(t,t)} {X(C)dC = \beta (1 -
\int_0^1 {C(X)dX)} }$ (in the simplest Ising case with
$C(t,t)=1$), formally equivalent to the Gibbs equilibrium
susceptibility in the Parisi replica symmetry breaking solution
for the Sherrington-Kirkpatrick model \cite{Mezard}, with
$C\Leftrightarrow q$ (overlap between pure states) and
$X\Leftrightarrow x$ (repartition of overlap). Theoretical
attempts, analytical \cite{Franz} (with the constraint of
stochastic stability) and numerical \cite{Marinari} (with the
problems of size effects), were made in order to confirm the above
properties in short range models. Up to now, experimental
investigations correspond only to the quasi-stationary regime
\cite{Grigera} or are very indirect \cite{CuGre}.

Here we report the result of an investigation of FD relation in
the insulating spin glass CdCr$_{1.7}$In$_{0.3}$S$_4$ \cite{Alba},
an already very well known compound, with $T_g=16.2K$. Above
$T_g$, the susceptibility follows a Curie-Weiss law $\chi  =
\mathcal{C}/(T -
\Theta ) $ where $ \mathcal{C}$ 
 corresponds
to ferromagnetic clusters of about 50 spins, and $\Theta\approx
-9K$ \cite{Vincent1}. The sample is a powder with grain sizes
around $10\, \mu m$, embedded in silicon grease to insure good
thermal contact between grains, and compacted into a coil foil
cylindrical sample holder $5\,mm$ wide and $40\,mm$ long. The two
times dependence of the magnetic relaxation (TRM) of this compound
was extensively studied \cite{Vincent2}.

\begin{figure}[b!]
\includegraphics[width=7cm]{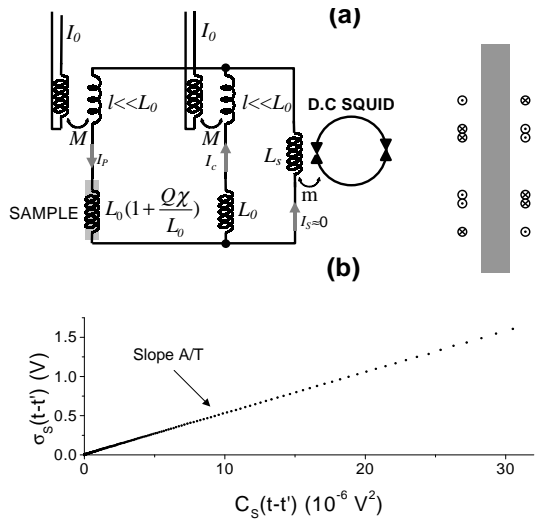}
\caption{\label{circuit} a) Schematic  of the detection circuit.
The pick-up coil (right side), containing the cylindrical sample,
is a third order gradiometer made of +3 -6 +6 -3 turns. b)
Calibration is obtained by measuring relaxation versus correlation
in a high conductivity copper sample at equilibrium at $4.2 K$.}
\end{figure}
In principle, SQUID measurement of magnetic fluctuations is very
simple \cite{Ocio,Refregier1}. The difficulty lies in the extreme
weakness of the thermodynamic fluctuations (of the order of the
response to a field about $10^{-7}\mathrm{G}$ in our case).
Therefore, the setup is carefully screened against stray fields by
superconducting shields, strict precautions are taken to suppress
spurious drifts of the SQUID electronics, and the pick-up coil is
a third order gradiometer. The result is that the proper noise
power spectrum of the system without sample allows time analysis
of the magnetic fluctuations signal over up to 2000 s of sample
fluctuations with more than 20 dB of signal/noise ratio. Moreover,
in the non-stationary regime, the time autocorrelation of magnetic
fluctuations
$C(t',t)=\frac{1}{N}\sum_i\left<\mathbf{m}_i(t')\mathbf{m}_i(t)\right>$,
where $\mathbf m_i$ is the elementary moment at site $i$, must be
determined as an ensemble average over a large number of records
of the fluctuation signal, each one initiated by a quench from
above $T_g$ (``birth" of the system). And finally, we want to
compare {\it quantitatively} correlation and relaxation data. The
relaxation function
$\sigma(t',t)=\frac{1}{N}\sum_i\left<\mathbf{m}_i(t)\right>/\mathbf{H}_i$
is measured by cooling the sample at time zero from above $T_g$ to
the working temperature in a small field, turning off the field at
time $t'$ and recording the magnetization at further times $t$.
Using a classical magnetometer with homogeneous field,
quantitative comparison between $C$ and $\sigma$ is almost
impossible due to the strong discrepancy between the coupling
factors in both experiments. Therefore, we have developed a new
bridge setup depicted in Fig.\ref{circuit}a, allowing measurements
of both fluctuations and response. The pick up (PU) coil of self
inductance $L_0$ is connected to the input coil of a SQUID, of
self inductance $L_S$. The whole circuit is superconducting.
Relaxation measurements use a small coil $l$ inserted in the
pick-up circuit, and coupled inductively with mutual inductance
$M$ to an excitation winding. A current $I_0$ injected in the
excitation results in a field induced by the PU coil itself
($\leqslant 1\mathrm{mG}$ here, clearly in the linear regime
though inhomogeneous), and the sample response is measured by the
SQUID. To get rid of the term $L_0$, the sample branch is balanced
by a similar one without sample, excited oppositely (see
Fig.\ref{circuit}a). The flux delivered to the PU by an elementary
moment $\mathbf{m}$ at position $\mathbf{r}$ is given by $\Phi (t)
= \mathbf{m}(\mathbf{r},t)\mathbf{h}(\mathbf{r}) $ where
$\mathbf{h}$ is the magnetic field produced by a unit of current
flowing in the PU. Flux conservation in the PU circuit results in
a current $I_S$ flowing in the input coil of the SQUID whose
output voltage is $V_S=GI_S$. Detailed analysis of the system will
be published elsewere. The main features are as follows.

As the fluctuations of elementary
moments in the sample are homogeneous and spatially uncorrelated
at the scale of the PU,
the SQUID output voltage autocorrelation is given by:
\begin{equation}
C_S(t',t)=\langle V_S (t')V_S (t)\rangle  = C(t',t)Q\frac{G^2} {{(L_0
+ 2L_S )^2 }}
\end{equation}
$Q = \sum_i {\mathbf{h}^2(\mathbf{r}_i )}$
where the index $i$ refers to a moment site,
is the coupling factor to the PU, including demagnetizing field
effects since $\mathbf{h}$ is the internal field.

The elementary moment response at site $i$ is $R_i (t',t) =
{\partial \left\langle {\mathbf{m}_i (t)} \right\rangle
}/{\partial \mathbf{h}(\mathbf{r}_i ,t')}$.Taking into account
that the medium is homogeneous, the relaxation function of the
SQUID output voltage is given by
\begin{equation}
\sigma_S(t',t)=\frac{V_S (t',t)}{I_0} =\sigma(t',t)Q\frac{MG}{L_0(L_0 + 2L_S) }.
\end{equation}
Thus, the coupling factor $Q$ disappears in the relation between
$C_S$ and $\sigma_S$, independently on the nature and shape of the
sample. There remains only the inductance terms $M$, $L_0$ and
$L_S$. These being difficult to determine with enough accuracy,
absolute calibration was performed using a copper sample of high
conductivity, by measuring $\sigma_S (t',t)$ and $C_S(t',t)$ ---
computed by standard fast Fourier transform algorithm ---  at
$4.2\mathrm{K}$ ($^4$He boiling temperature at normal pressure):
with this ergodic material, the relation between both measured
quantities is linear with slope $A/T$, where $A$ is the
\textbf{\textit{ sample independent}} calibration factor (see
Fig.\ref{circuit}b). From the knowledge of $A$, determined at
$4.2\mathrm{K}$, the system is equivalent to a thermometer, i.e.
the FDT slope is known \textbf{\textit{exactly}} at any
temperature.

\begin{figure}[b!]
  \includegraphics[width=7cm]{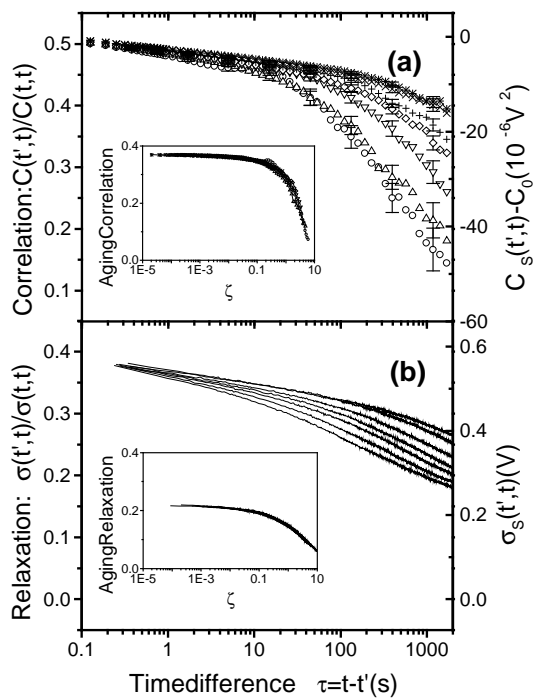}
  \caption{\label{figure2}Aging and scaling of (a) correlation
(b) relaxation at $T=0.8T_g$. Both are measured for waiting times
$t'=100$ ($\circ$), 200 ($\triangle$) , 500($\triangledown$),
1000($\diamond$), 2000 ($+$) , 5000($\times$), 10000($*$) seconds
from bottom to top. Reported error-bars on correlation have a
length of two standard-deviation, corresponding to averages over
records. In insets, scaling of the aging parts versus $\zeta=(
t^{1-\mu} - t'^{1-\mu} )/(1-\mu)$, using $\mu=0.87$. The
stationary parts are found to obey a power-law decrease with an
exponent $\alpha=0.05$.}
\end{figure}

In the spin glass sample, $C_S(t',t)$ and $\sigma_S(t',t)$ were
measured at $T=0.8T_g$ after quench from a temperature $T\approx
1.2T_g$. To get a precise definition of the ``birth" time, a
minimum value of 100 s was chosen for $t'$. The autocorrelation
was determined from an ensemble of 320 records of
$12000\,\mathrm{s}$ of the fluctuation signal. The ensemble
averages were computed in each record from the signal at $t'$,
averaged over $\delta t' \leqslant t'/20$, and the one at $t$,
averaged over $\delta t \leqslant (t - t')/10$ --- the best
compromise allowing a good average convergence still being
compatible with the non-stationarity --- , and averaging over all
records. As there is an arbitrary offset in the SQUID signal, the
connected correlation was computed. Nevertheless, this was not
enough to suppress the effect of spurious fluctuation modes of
period much longer than 2000 s, giving a non-zero average offset
on the correlation results. Thus, as a first step, we have plotted
all correlation data, taking as \textbf{\textit{the}} origin the
value of $\langle V_S^2(t')\rangle$. Due to the elementary
measurement time constant this last term corresponds to an average
over $t-t'$ about $10^{-2}$s, i.e. a range of $(t-t')/t'$
corresponding to stationary regime. Thus, all $C_S$ data are
shifted by a common offset $C_0$. The result is shown in
Fig.\ref{figure2}a (right sided scale), as a function of $t-t'$
for values of $t'$ from 100 s to 10000 s. Residual oscillations
---and large error bars--- for $t'=100s$ reveal the limit of
efficiency of our averaging procedure. Corresponding relaxation
data are plotted on Fig.\ref{figure2}b. In both results, one can
see that the curves merge at low $t-t'$ , meaning that they do not
depend on $t'$ (stationary regime). At $t-t'\geqslant t'$, they
strongly depend on $t'$, the slower decay corresponding to the
longer $t'$.

\begin{figure}
  \includegraphics[width=8cm]{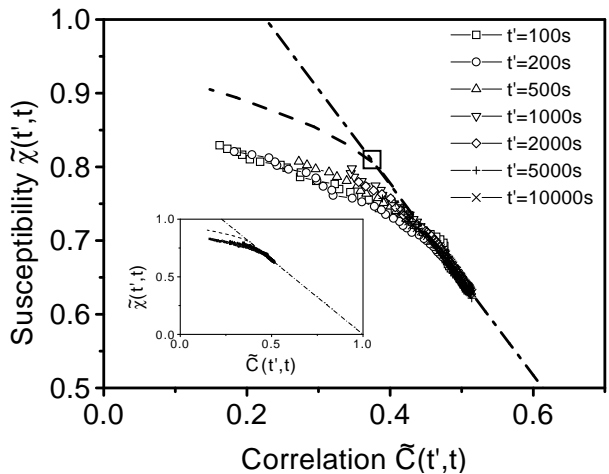}
  \caption{\label{figure3}FD-plot. Relaxation measurements are plotted
versus correlation functions for each $t'$. The dot-dashed line
(FDT line) is calculated for $T= 0.8Tg = 13.3 K$, from the
calibration obtained with the copper sample. The dashed line represents the scaling
extrapolation for $t'\rightarrow\infty$. The branching point with
the FDT line, corresponds to {$\widetilde{C}=q_{EA}$} (square
symbol, with size giving the error range). In Inset, the same data
in the whole range.}
\end{figure}

The correlation offset must be determined. As zero of correlation
is unreachable in experimental time, correction of the offset
could be obtained from the knowledge of $C(t,t)$. Nevertheless,
due to clustering, $C(t,t)$ depends on temperature and cannot be
determined from the high temperature susceptibility. In canonical
compounds like 1\% Cu:Mn \cite{Binder}, with negligible
clustering, the field-cooled susceptibility is temperature
independent in agreement with the Parisi-Toulouse hypothesis
\cite{Marinari,Parisi}, yielding $C(t,t)=T_g \chi_{FC}(T)$. We
used a generalization of this relation with the condition that a
smooth dependence of $C(t,t;T)/T$ must result \cite{Parisi2}. This
was obtained by using for $T_g$ a slightly different value,
$T_g^*=17.2 \mathrm{K}$. Then, from the value of the calibration
factor $A$, and writing $C(t,t;0.8T_g)=17.2\chi_{FC}(0.8T_g)$,
$C_S(t,t;0.8T_g)$ can be determined, and suppression of the offset
can be performed by using the $\chi (C)$ plot, first introduced by
Cugliandolo and Kurchan \cite{CuKu2}. We plot the normalized
susceptibility function $\widetilde{\chi}(t',t)=1
-\widetilde\sigma(t',t)$ where
$\widetilde\sigma(t',t)=\sigma_S(t',t)/\sigma_S(t,t)$ (note that
$\sigma(t,t)=\chi_{FC}$) versus normalized autocorrelation
$\widetilde C(t',t)-\widetilde C_0=(C_S(t',t)-C_0)/C_S(t,t;T)$ for
all experimental values of $t'$. In this graph, the FDT line has
slope $-T_g^*/T$ and crosses the $\widetilde C$ axis at
$\widetilde C=1$. On the data, a clear linear range appears at
large $\widetilde{C}$ (small $t-t'$), displaying the FDT slope
with error $<3\%$ in the sector $\widetilde C\geqslant 0.47$. This
allows the suppression of the correlation offset by horizontal
shift of the data. The result is shown in Fig.\ref{figure3}. It is
of course based on a rough ansatz on $C(t',t;T)$ which needs
further justifications, but we stress that the induced uncertainty
concerns only the position of the zero on the $\widetilde C$ axis,
and not the shape and slope of the curves. With decreasing
$\widetilde{C}$ (increasing $t-t'\geqslant t'$), the data points
depart from the FDT line. Indeed,
$\widetilde{C}(t',t\rightarrow\infty)=0$ and
$\widetilde{\chi}_{FC}=\widetilde{\chi}(t',t\rightarrow\infty)
=1$. The mean slope of the off FDT data corresponds to a
temperature of about $30 \mathrm{K}$. This value is far above our
annealing temperature, ruling out a simple interpretation in terms
of a ``fictive" temperature \cite{Jackle}. Despite the scatter of
the results, a tendency for the data at small $t'$ to depart the
FDT line at larger values of $\widetilde{C}$ is clear: it is
experimentally impossible to fulfill the condition of timescales
separation underlying the existing theories . Even if the long
$t'$ limit for $\widetilde{\chi}(\widetilde{C})$ does exist, it is
not reached in the plot of data in Fig.\ref{figure3} and a $t'$
dependence of the $\widetilde{\chi}(\widetilde{C})$ curves is
expected.

The left sided scales in Fig.\ref{figure2}a and b correspond to
$\widetilde C(t',t)$ and $\widetilde\sigma(t',t)$ respectively. In
former works, it was shown that the whole relaxation curves could
be scaled as the sum of two contributions, one stationary and one
non-stationary \cite{Vincent2}
\begin{equation}
\widetilde\sigma(t',t) = (1 - \Delta )(1 + (t - t')/t_0 )^{ - \alpha
}  + \Delta \varphi (\zeta ),
\end{equation}
where $t_0$ is an elementary time of order $10^{-11}$ s, $\varphi$
is a scaling function of an effective time parameter $\zeta
\propto t^{1 - \mu }  - t'^{1 - \mu }$ depending on the sub-aging
coefficient $\mu<1$ \cite{Vincent2}, and $\alpha$ can be
determined with good precision from the stationary power spectrum
of fluctuations $S(\omega)\propto\omega^{\alpha-1}$. The inset in
Fig.\ref{figure2}b displays the result of the scaling on the
relaxation curves with $\alpha=0.05$, $\Delta=0.21$ and
$\mu=0.87$. As shown in the inset of Fig.\ref{figure2}a, the
scaling works rather well on the autocorrelation curves with the
same exponents, but now, $q_{EA}$, the Edwards Anderson order
parameter, replaces $\Delta$. We get $q_{EA}=0.37$. These results
show clearly that the stationary part of the dynamics is still
important yet in the aging regime, i.e. that the limit of long
$t'$ is not reached within the timescale of our experiments (in
fact, timescale separation is realized if $t'\geqslant \tau$ where
$\tau$ is the observation time such that $C_{stat} (\tau)\ll
q_{EA}$).

If granted, the scaling gives the long time limit of the
non-stationary part of the dynamics, allowing a plot of the long
times asymptotic non-stationary part of the
$\widetilde{\chi}(\widetilde{C})$ curve. Of course, here we verify
it only over 2 decades of time, up to $t'=10000\,\mathrm{s}$, but
it was proven to be relevant on TRM up to $t'=100000\,\mathrm{s}$
\cite{Alba}. The dashed line in Fig.\ref{figure3} is obtained by
plotting the smoothed curves of aging parts of
$\widetilde{\chi}(\zeta)$ versus $\widetilde{C}(\zeta)$. According
to theoretical conjectures,
$d\widetilde{\chi}(\widetilde{C})/d\widetilde C$ would represent
the static quantity $x(q)$ \cite{Franz}. One can see that the
curve does not point exactly towards $\widetilde{\chi}=1$ but
about 5\% below. Therefore i) either the ansatz used to determine
$C(t,t;T)$ is not realistic enough, ii) or the time scaling is no
longer valid at the very large $t'$ needed for timescales
separation. For future progress, scaling developments outside the
strict time range separations at the basis of the ``adiabatic
cooling" analysis \cite{Dotsenko} or the ``weak memory" analysis
\cite{CuKu3} are needed. It seems that such developments are out
of the theoretical possibilities for the moment. Toy models
\cite{Ocio2} presently under development could allow a
phenomenological approach of the problem.

In conclusion, we have presented the first experimental
determination of the non stationary time autocorrelation of magnetization in a
spin glass, an archetype of a complex system. With the help of the
time scaling properties of both the relaxation and the
autocorrelation, we were able to propose a first experimental
approach of a possible generalization of FDT to non-stationary
systems. Results at several temperatures are now needed in order to
get a complete description of the $\widetilde{\chi}(\widetilde{C})$
behavior in the whole temperature range.

We thank J. Hammann, E. Vincent, V. Dupuis, L. F. Cugliandolo, J. Kurchan, D.
R. Grempel, M. V. Feigel'man, L. B. Ioffe and particularly G. Parisi for
enlightening discussions and
critical reading of the manuscript. We are indebted to P. Monod
for providing the high conductivity copper sample.

\end{document}